\documentclass[prb,twocolumn,showpacs,showkeys,preprintnumbers,amsmath,amssymb]{revtex4}

\usepackage{bm} 
\usepackage{graphicx} 
\newcommand{\df}{$\Delta F$}
\newcommand{\dfrs}{$\Delta F_{\rm ref \rightarrow phys}$}

\begin{document}

\title{Simple estimation of absolute free energies for biomolecules}
\author{ F.\ Marty Ytreberg\footnote{E-mail: fmy1@pitt.edu} }
\author{ Daniel M.\ Zuckerman\footnote{E-mail: dmz@ccbb.pitt.edu} }
\affiliation{Department of Computational Biology,
  School of Medicine, University of Pittsburgh, Pittsburgh, PA 15261}
\date{\today}

\begin{abstract}
One reason that free energy difference calculations are notoriously
difficult in molecular systems
is due to insufficient conformational overlap, or similarity, between
the two states or systems of interest. The degree of overlap is irrelevant,
however, if the absolute free energy of each state can be computed. We present
a method for calculating the absolute free energy
that employs a simple construction of an exactly computable
reference system which
possesses high overlap with the state of interest. The approach
requires only a physical ensemble of conformations generated via
simulation, and an auxiliary
calculation of approximately equal central-processing-unit (CPU) cost.
Moreover, the calculations can converge to the correct free energy value
even when the physical ensemble is incomplete or improperly distributed.
As a ``proof of principle,''
we use the approach to correctly predict free energies for
test systems where the absolute values can be calculated
exactly, and also to predict the
conformational equilibrium for leucine dipeptide in 
implicit solvent.
\end{abstract}
\keywords{free energy,entropy}
\pacs{pacs}
\maketitle

\section{Introduction}
Knowledge of the free energy for two different
states or systems of interest
allows the calculation of solubilities,
\cite{grossfield-jacs,vangunsteren-onestep}
determines binding affinities of ligands to proteins,
\cite{kollman-pnas,vangunsteren-estrogen}
and determines conformational equilibria
(e.g., Ref.\ \onlinecite{ytreberg-shift}).
Free energy differences (\df) therefore have potential
application in structure-based drug design where current
methods rely on {\it ad hoc} protocols to estimate binding affinities.
\cite{shoichet-nature,scheraga}

Poor ``overlap,'' the lack of configurational
similarity between the two states or systems of interest,
is a key cause of computational expense and error in \df\ calculations.
The most common approach to improve overlap in free energy
calculations (used in thermodynamic integration, and free energy
perturbation) is
to simulate the system at multiple hybrid, or intermediate stages
(e.g., Refs.\ \onlinecite{zwanzig,beveridge,jorgensen,karplus-jcp,mccammon}).
However, the simulation of intermediate stages
greatly increases the computational cost of the \df\ calculation.

Here, we address the overlap problem by calculating the absolute free
energy for each of the end states, thus avoiding the need for any
configurational overlap. Our method relies on the calculation of
the free energy difference between
a reference system (where the exact free energy
can be calculated, either analytically or numerically)
and the system of interest.

Such use of a reference system with a computable free energy
has been used successfully in solids
where the reference system is generally a harmonic or
Einstein solid, \cite{hoover71,frenkel}
and liquid systems, where the reference
system is usually an ideal gas. \cite{hoover67,reinhardt-absf}
The scheme has also been applied to molecular
systems by Stoessel and Nowak, using a harmonic
solid in Cartesian coordinates as a reference system. \cite{stoessel}

Other approaches to calculate the absolute free energies of
molecules have been developed.
Meirovitch and collaborators calculated
absolute free energies for peptides in vacuum,
for liquid argon and water using the hypothetical
scanning method. \cite{meirovitch-deca,meirovitch-argon}
Computational cost has thus far limited the approach to
peptides with sixty degrees of freedom. \cite{meirovitch-jcp}
The ``mining minima'' approach, developed by Gilson and collaborators,
estimates the absolute free energy of complex molecules
by attempting to enumerate the low-energy conformations 
and estimating the contribution to the configurational integral
for each. \cite{gilson-jpca,gilson-bj}
Anharmonic effects can be included. \cite{gilson-jacs}
The mining minima method can, in principle, include potential
correlations between the torsions and bond angles or lengths,
and uses an approximate method to compute local partition functions.
Other investigators have estimated absolute free energies for molecules
using harmonic or quasi-harmonic approximations,
\cite{karplus-deca,gilson-jacs,aqvist-absf}
however, as discussed in
Refs.\ \onlinecite{gilson-jacs} and \onlinecite{karplus-deca}
local minima can be deviate substantially
from a parabolic shape.

We introduce, apparently for the first time, a reference system
which is constructed to have high overlap with fairly general
molecular systems. The approach
can make use of either {\it internal or Cartesian}
coordinates. For biomolecules, using internal coordinates greatly  
enhances the accuracy of the method since internal coordinates
are tailored to the description of conformations.
Further, {\it all degrees of freedom and their correlations}
are explicitly included in the method.

Our method differs in several ways from the important study of
Stoessel and Nowak: \cite{stoessel}
(i) we use internal coordinates
for molecules which are key for optimizing the overlap between
the reference system and the system of interest;
(ii) we may use a nearly arbitrary reference potential because
only a numerical reference free energy value is needed,
not an analytic value;
(iii) there is no need, in cases we have studied,
to use multi-stage methodology to find
the desired free energy due to the overlap built into the
reference system, 

We consider this report a ``proof of principle''
for our reference system method.
After introducing the method,
it is tested on single and double-well two-dimensional systems,
and on a methane molecule where absolute free energy 
estimates can be compared to exact values.
The method is then used to compute the absolute free energy
of the alpha and beta conformations for
leucine dipeptide (ACE-(leu)$_2$-NME) in implicit solvent,
{\it using all one-hundred fifteen degrees of freedom},
correctly calculating the free energy difference 
$\Delta F_{\rm alpha \rightarrow beta}$.
Extensions of the method to larger systems are
then discussed.

\section{Reference system method\label{sec-method}}
\subsection{The fundamental relations}
The absolute free energy of the system of interest (``phys'' for physical)
is defined using the partition function $Z_{\rm phys}$
\begin{eqnarray}
    F_{\rm phys} = -k_BT \ln Z_{\rm phys} = \nonumber \\ 
	-k_BT \ln \left[ 
	    \int d \vec{x} \; 
	    e^{-\beta \big(U_{\rm phys}(\vec{x})+K_{\rm phys}(\vec{x})\big)}
	\right],
\end{eqnarray}
where $T$ is the system temperature, $\beta=1/k_BT$,
$U_{\rm phys}$ and $K_{\rm phys}$ are, respectively, the
physical potential energy (i.e., simulation forcefield)
and the kinetic energy,
and $\vec{x}$ represents the full set of
configurational coordinates (internal or Cartesian).
The kinetic energy term can be integrated exactly to obtain
\cite{gilson-jpcb}
\begin{eqnarray}
    Z_{\rm phys} =
	\Bigg[ \frac{1}{h^{3N}}\frac{8\pi^2}{\sigma C^{\circ}}
	\prod_{i=1}^N \big( 2\pi k_B T m_i \big)^{3/2} \Bigg] 
	    \int d \vec{x} \; e^{-\beta U_{\rm phys}(\vec{x})},
    \label{eq-Zphys}
\end{eqnarray}
where  $m_i$ is the mass of atom $i$, $h$ is Planck's constant,
$C^{\circ}$ is the standard concentration,
$\sigma$ is the symmetry number, \cite{gilson-bj}
$N$ is the number of particles in the system,
and the integral is
defined to be the configurational partition function.
For method used in this study the absolute free energy of the system
of interest is calculated using
a reference system (``ref''), and the following relationships are used,
\begin{eqnarray}
    Z_{\rm phys} = Z_{\rm ref} \frac{Z_{\rm phys}}{Z_{\rm ref}}, \nonumber \\
    F_{\rm phys} = F_{\rm ref} + \Delta F_{\rm ref \rightarrow phys},
    \label{eq-Fphys}
\end{eqnarray}
where $F_{\rm ref}$ is the trivially computable
free energy of the reference system, and \dfrs\ is the free energy
difference between the reference and physical system which can
be calculated using standard techniques.

For this report, we include estimates of the configurational
integral only, i.e., the leading constant factor in square brackets in
Eq.\ (\ref{eq-Zphys}) is not included in our results. Ignoring
the constant is not a limitation since, for the conformational free
energies studied here, the term cancels for
free energy differences.

\subsection{The reference energy and its normalization}
\begin{figure}
    \includegraphics[scale=0.35]{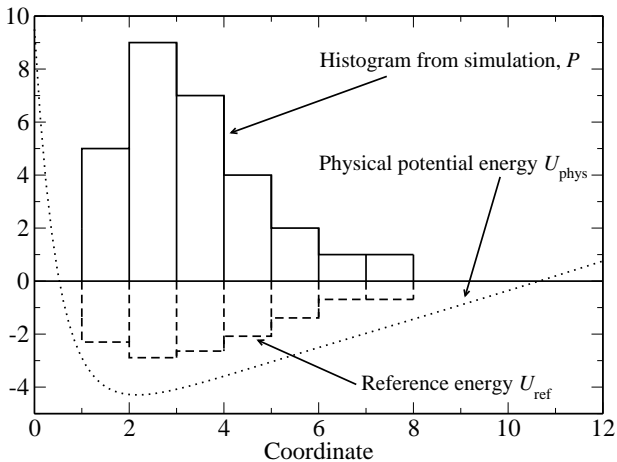}
    \caption{Depiction of how the reference potential energy
	$U_{\rm ref}$ is calculated for a one-coordinate system.
	First the coordinate is binned, creating a 
	histogram $P$ (solid bars) populated according to the physical
	ensemble. Then Eq.\ (\ref{eq-Uref}) is used to
	calculate reference energies for each coordinate bin (dashed bars).
	A hypothetical physical potential is
	shown as a dotted curve for comparison to $U_{\rm ref}$.
	For a multi-coordinate system $U_{\rm ref}$
	would be the sum of the single-coordinate reference
	potential energies.
	\label{fig-schematic}
    }
\end{figure}

The trivial identities of Eq.\ (\ref{eq-Fphys}) suggest that arbitrary
reference systems can be used in our approach. To be concrete and anticipate
the procedure used, our discussion below will assume that a finite-length
simulation of the system of interest has been performed---from which
histograms of the coordinates have been generated.
For the molecular systems studied in this report, ordinary
Langevin dynamics simulations are performed using standard
forcefields.
The reference potential energy can be constructed from a wide
variety of histograms, as discussed below. Denoting
the computed histograms over all coordinates as $P(\vec{x})$, we define
\begin{eqnarray}
    U_{\rm ref}(\vec{x}) \equiv -k_BT \ln P(\vec{x}),
    \label{eq-Uref}
\end{eqnarray}
where $P(\vec{x})$ is the normalized probability of a particular
configuration (corresponding to a set of histogram bins);
see Fig.\ \ref{fig-schematic}.
For example, if all coordinates are binned as independent, then
\begin{eqnarray}
    P(\vec{x})=\prod_{i=1}^{N_{\rm coords}} P_i(x_i),
    \label{eq-Pind}
\end{eqnarray}
where $P_i(x_i)$ is the binned probability distribution (histogram)
for the $i^{\rm th}$ coordinate, and there
are $N_{\rm coords}$ degrees of freedom in the system.
If all coordinates are binned as pairwise correlated, then
\begin{eqnarray}
    P(\vec{x})=\prod_{ \{ i,j \} } P_{ij}(x_i,x_j),
    \label{eq-Pcorr}
\end{eqnarray}
where $\{ i,j \}$ is a set of pairs in which each coordinate occurs exactly
once, and $P_{ij}(x_i,x_j)$ is the probability for two particular coordinate
values from the two-dimensional histogram for these coordinates.
It is also possible to use an arbitrary combination of independent
and correlated coordinates---so long as each coordinate occurs
in only one $P$ factor.

We emphasize that the final computed free energy values include
all correlations embodied in the true potential $U_{\rm phys}$. This
is true regardless of whether or how coordinates are correlated in the
reference potential.

A schematic of how $U_{\rm ref}$ is computed
for a one-coordinate system is shown in Fig.\ \ref{fig-schematic}.
The coordinate histogram is first determined (solid bars)
using a simulation trajectory;
then Eq.\ (\ref{eq-Uref}) is used to calculate
$U_{\rm ref}$ (dashed bars). A possible physical potential is
also included (dotted line) for comparison to $U_{\rm ref}$.
For a system containing many degrees of freedom,
the process is carried out for all coordinates, based
on Eq.\ (\ref{eq-Pind}), (\ref{eq-Pcorr}) or other correlation scheme.
$U_{\rm ref}$ is the sum
of all the appropriate terms,
consistent with Eq.\ (\ref{eq-Uref}) and the binning choice.

The free energy of the reference system can now
be calculated via the reference partition function
\begin{eqnarray}
    Z_{\rm ref} = \int d\vec{x} \; e^{-\beta U_{\rm ref}(\vec{x})}
        = \int d\vec{x} \; P(\vec{x}). 
    \label{eq-Zref}
\end{eqnarray}
In practice, we normalize the histogram for each coordinate
to one independently by summing
over all histogram bins. So, for a particular bond length $r_1$,
that is binned as independent, we account for the Jacobian
factor (see Eq.\ (\ref{eq-jacobian})) by defining $\xi = r_1^3/3$, and then
\begin{eqnarray}
    Z_{\xi} = \int d\xi \; P(\xi)
	= \sum_{N_{\rm bin}} \; \Delta \xi \; P(\xi) = 1, 
\end{eqnarray}
where $\Delta \xi$ is the histogram bin size, and $N_{\rm bin}$
is the number of bins in the $r_1$ histogram.
(Binning choices are discussed below.)
Similar relationships are used for all coordinates.
Thus the reference free energy $F_{\rm ref}=0$
and Eq.\ (\ref{eq-Fphys}) becomes 
\begin{eqnarray}
    F_{\rm phys} = \Delta F_{\rm ref \rightarrow phys}
    \;\;\;\;\;\; (F_{\rm ref} \equiv 0)
    \label{eq-Fphys2}
\end{eqnarray}

\subsection{Using the physical and reference ensembles}
With the reference potential energy $U_{\rm ref}$
defined in Eq.\ (\ref{eq-Uref})
and the physical potential energy $U_{\rm phys}$
given by the forcefield, which may include implicit solvation energies,
Boltzmann-distributed snapshots from both the
reference and physical systems can be utilized
to calculate $F_{\rm phys}$=\dfrs.
Here, we simply use free energy perturbation \cite{zwanzig}
from the reference to the physical systems
\begin{eqnarray}
    F_{\rm phys} = -k_B T \ln \Big\langle
	e^{-\beta \big( U_{\rm phys}-U_{\rm ref} \big) }
    \Big\rangle_{\rm ref} \nonumber \\
        \doteq -k_B T \ln \Bigg( \frac{1}{N_{\rm ref}} \sum_{i=1}^{N_{\rm ref}}
	e^{-\beta \big(U_{\rm phys}-U_{\rm ref}\bigr)} \Bigg)
    \label{eq-fep}
\end{eqnarray}
where $N_{\rm ref}$ is number of structures in the reference ensemble,
the ``$\doteq$'' symbol denotes a computational estimate,
and $\langle ... \rangle_{\rm ref}$ represents a canonical average
using structures from the reference ensemble only.
It is important to note that, while other choices for
computing $F_{\rm phys}$ are possible, such as Bennett's method,
\cite{bennett,shirts-benn,shirts-prl,crooks-pre,lu-jcc,ytreberg-shift}
Eq.\ (\ref{eq-fep}) is the only choice which relies solely on
configurations drawn from the reference ensemble which
are, by construction, sampled canonically and without
dynamical trapping.
We also note that ``uni-directional'' estimates like that of
Eq.\ (\ref{eq-fep}) have been analyzed extensively
(e.g., Refs.\ \onlinecite{zuckerman-prl} and \onlinecite{zuckerman-jstat})
and may be amenable to error-reduction techniques;
\cite{zuckerman-cpl,ytreberg-extrap} however, we have applied the
perturbation approach here to keep our initial analysis as straightforward
as possible.
Staged free energy methods like thermodynamic
integration \cite{straatsma-ti} and adaptive integration
\cite{swendsen-aim} may also be used.

\subsection{The physical ensemble and construction of the reference system}
The method used in this report relies on simple
histograms for all degrees of freedom
(in principle, with internal or Cartesian coordinates)
based on a ``physical ensemble'' of
conformations generated via molecular dynamics,
Monte Carlo or other canonical simulation.
The histograms define a reference system with a free energy that is
trivially computable, as described in Sec.\ \ref{sec-method}.
We emphasize that an analytical
solution need not be available; a precise numerical evaluation is
more than adequate.
A well-sampled ensemble of reference system configurations is then
readily generated and used to compute the free energy
difference via Eq.\ (\ref{eq-fep}).

The first step in our approach to constructing the reference
system is to generate a physical
ensemble (i.e., a trajectory) by simulating the
system of interest using
standard molecular dynamics, Monte Carlo, or other
canonical sampling techniques.
The trajectory produced by the simulation is
used to generate histograms for all coordinates
as described below.
In creating histograms, note that constrained coordinates,
such as bond lengths involving hydrogens constrained
by RATTLE, \cite{rattle}
need not be binned since these coordinates do not change
between configurations.
Such coordinate constraints are not required in the method, however.

If internal coordinates are used (such as for the molecules
in this study), care must be taken to
account for the Jacobian factors.
Using internal coordinates with bond lengths $r$,
bond angles $\theta$ and dihedrals $\omega$, the
volume element in the configurational integral
of Eq.\ (\ref{eq-Zphys}) is given by \cite{gilson-jacs}
\begin{eqnarray}
    d \vec{x} = 
	\prod_{i=1}^{N-1} r_i^2 dr_i \;
	\prod_{i=1}^{N-2} \sin\theta_i d\theta_i \; 
	\prod_{i=1}^{N-3} d\omega_i
    = \nonumber \\
	\prod_{i=1}^{N-1} d (r_i^3/3) \;
	\prod_{i=1}^{N-2} d (-\cos \theta_i) \; 
	\prod_{i=1}^{N-3} d\omega_i, 
    \label{eq-jacobian}
\end{eqnarray}
where $N$ is the number of atoms in the system.
Thus, when using internal coordinates,
the simplest strategy to account for the
Jacobian is to bin according to a set of rules:
bond lengths are binned according to $r^3/3$,
bond angles are binned according to $\cos\theta$,
and dihedrals are binned according to $\omega$ (i.e., the
same as Cartesian coordinates).

\subsection{Generation of the reference ensemble}
Once the histograms are constructed and populated using the physical
ensemble, the reference ensemble is generated.
To generate a single reference structure,
for each coordinate one chooses a histogram
bin according to the probability associated with that bin. Then a
coordinate value is chosen at random uniformly
within the bin according
the Jacobian factor in Eq.\ (\ref{eq-jacobian})---e.g., for
a bond length $r$, one chooses uniformly in the variable $(r^3/3)$.
The process is repeated for every degree of freedom in the system.
By repeating the entire procedure, one can generate
as many reference structures as desired
(i.e., the reference ensemble).

\subsection{Summary of the reference system method}
In summary, the method is implemented by first constructing
properly normalized histograms for all internal (or Cartesian) coordinates
based on a physical ensemble of structures.
An ensemble of reference structures is then chosen at random from the
histograms.
The reference energy ($U_{\rm ref}$ of Eq.\ (\ref{eq-Uref})) and
physical energy ($U_{\rm phys}$ from the forcefield) must
be calculated for each structure in the reference ensemble.
Finally, Eq.\ (\ref{eq-fep}) is used to calculate the
desired absolute free energy of the system of interest.

The CPU cost of the method, above that of the
initial ``physical'' trajectory, is one physical energy evaluation
for each of the $N_{\rm ref}$ reference structures, plus the less
expensive cost of generating reference structures.

\section{Results}
To test the effectiveness of the reference system method
we first estimated the absolute free energy for three test systems
where the free energy is known exactly.
We chose the two-dimensional potentials
from Ref.\ \onlinecite{ytreberg-seps}, and  a methane molecule in vacuum.
Finally, we used the method to estimate the absolute free energies
of the alpha and beta conformations of the 50-atom
leucine dipeptide (ACE-(leu)$_2$-NME), and compared
the free energy difference obtained via our method
with an independent estimate.
In all cases, the free energy estimate computed by our approach
is in excellent agreement with independent results.

\subsection{Simple test systems}
We first studied the two-dimensional
single and double-well potentials from Ref.\ \onlinecite{ytreberg-seps},
\begin{eqnarray}
  U_{\rm phys}^{\rm single}(x,y)=(x+2)^2+y^2, \nonumber\\
  U_{\rm phys}^{\rm double}(x,y)=\frac{1}{10}
  \Bigl\{
  ((x-1)^2-y^2)^2+ \nonumber \\
  10(x^2-5)^2 + (x+y)^4+(x-y)^4
  \Bigr\}.
  \label{eq-pot}
\end{eqnarray}

\begin{table}
    \begin{tabular}{l|c|c}
    \hline \hline
    System & Exact & Estimate \\
    \hline
    two-dimensional single-well \cite{ytreberg-seps} & -1.1443 & -1.1449 (0.0003) \\
    \hline
    two-dimensional double-well \cite{ytreberg-seps} & 5.4043 & 5.4058 (0.0003)\\
    \hline
    Methane molecule & 10.932 & 10.934 (0.002)\\
    \hline \hline
    \end{tabular}
    \caption{
	Absolute free energy estimates obtained using our 
	reference system approach for cases where the absolute free
	energy can be determined exactly.
	In all cases, the estimate is in excellent agreement with
	the exact free energy.
	The uncertainty, shown in parentheses
	(e.g., $3.14 \; (0.05) = 3.14 \pm 0.05$), is
	the standard deviation from five independent simulations.
	The results for the two-dimensional systems are in $k_BT$ units
	and methane results have units of kcal/mole.
	The table shows estimates of the configurational
	integral in Eq.\ (\ref{eq-Zphys}),
	i.e., the constant term is not included in the estimate.
    \label{tab-results}
    }
\end{table}

Table \ref{tab-results} shows the excellent agreement
between the reference system estimates and the exact free energies
(obtained analytically) for the
two-dimensional potentials used in this study, Eq.\ (\ref{eq-pot}).
The ``physical'' simulations used Metropolis Monte Carlo
with $k_BT=1.0$ and one
million snapshots in the physical and reference ensembles.
For all two-dimensional simulations, both coordinates
were treated with full
correlations---i.e., two-dimensional histograms were used---and
the bin sizes were chosen such that the number of bins ranged from
100-1000.
The error shown in Table \ref{tab-results}
in parentheses is the standard deviation from five independent estimates
using five separate physical ensembles---and thus five different
reference systems.
Good estimates were also obtained using fewer snapshots---e.g.,
we obtained $F=-1.142 \; (0.003)$
for the single-well potential
and $F=5.408 \; (0.007)$ for the double-well potential
using 10,000 snapshots
in both the physical and reference ensembles.

Table \ref{tab-results} also shows the excellent agreement between the
reference system estimates and the exact value of the free
energy for methane in vacuum.
Methane trajectories were generated
using TINKER 4.2 \cite{tinker} with the OPLS-AA forcefield. \cite{oplsaa}
The temperature was maintained at 300.0 K using Langevin dynamics with
a friction coefficient of 91.0 ${\rm ps}^{-1}$ and a time step of 0.5 fs.
The physical ensemble was created by generating five 10.0 ns trajectories 
with snapshots saved every 0.1 ps.
Using the 100,000 methane structures in the physical ensemble,
the reference system was generated by binning internal coordinates
into histograms. The absolute free energy was then estimated
by generating 100,000 structures for the reference ensemble
and using Eq.\ (\ref{eq-fep}).
All coordinates were binned as independent using
one-hundred bins per coordinate, thus only one-dimensional histograms
were required.
The uncertainty shown in parenthesis in Table \ref{tab-results}
is the standard deviation from
five independent estimates using the five separate methane
trajectories---and thus five different reference systems.

\begin{figure}
    \includegraphics[scale=0.35]{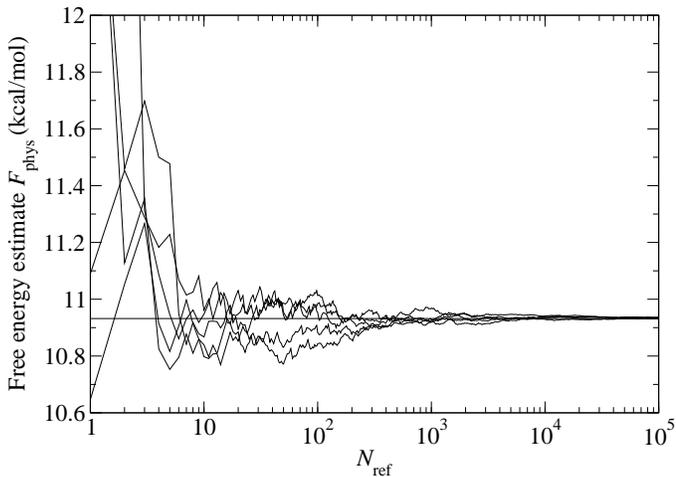}
    \caption{Absolute free energy for methane estimated by
	the reference system
	method as a function of the number of reference
	structures $N_{\rm ref}$ used in the estimate.
	The solid horizontal line 
	is the exact free energy obtained by numerical integration.
	Five independent simulations are shown on a log scale to clearly
	show the convergence of the free energy estimate.
	Results shown were obtained using Eq.\ (\ref{eq-fep})
	with one-hundred bins for each degree of freedom, i.e., the estimates
	for the absolute free energy of methane in Table \ref{tab-results}
	are the values shown here for
	$N_{\rm ref}=1,000,000$.
	\label{fig-converge-meth}
    }
\end{figure}

Figure \ref{fig-converge-meth} shows the convergence
behavior of the reference
system method for methane. Five independent absolute free energy
estimates are shown as a function of the number of reference
structures used in the estimate.
Each of the five simulations use the same protocol as described above,
i.e., the absolute free energy estimates in Table \ref{tab-results} are
the values shown in
Fig.\ \ref{fig-converge-meth} for $N_{\rm ref}=100,000$.

Methane was chosen as a test system because
intra-molecular interactions are due only to bond
lengths and angles. In the OPLS-AA forcefield no non-bonded terms
are present in the
potential energy $U_{\rm phys}$, and thus the exact absolute free energy can
be computed numerically without great difficulty.
For methane, a configuration is determined by:
(i) four bond lengths, which are independent of each other and 
all of other coordinates in the forcefield; and
(ii) five bond angles which are correlated to one another but
not to the bond lengths.
Thus the exact partition function $Z_{\rm meth}$ is a product
of four bond length partition functions $Z_r$ and one
angular partition function $Z_{\theta}$,
\begin{eqnarray}
    Z_{\rm meth} = Z_r^4 Z_{\theta}, \nonumber \\
	Z_r = \int_{0}^{\infty} dr\;e^{-\beta U_{\rm phys}(r)},
	    \nonumber \\
	Z_{\theta} = \int_{0}^{\pi}
	    d\theta_1 d\theta_2 d\theta_3 d\theta_4 d\theta_5 \;
	    e^{-\beta U_{\rm phys}
		(\theta_1,\theta_2,\theta_3,\theta_4,\theta_5)
	      }.
\end{eqnarray}
$U_{\rm phys}(r)$ is harmonic and thus $Z_r$ was computed analytically
using parameters from the forcefield.
For $U_{\rm phys}(\theta_1,\theta_2,\theta_3,\theta_4,\theta_5)$
the correlations between angles must be
taken into account, thus $Z_{\theta}$ was estimated numerically using
TINKER to evaluate $U_{\rm phys}$ in the five-dimensional integral.
We found that $F_{\rm meth}=-k_B T \ln Z_{\rm meth} = 10.932$ kcal/mol
as shown in Table \ref{tab-results}.

Methane was also used to show that the method correctly computes
the free energy even when the physical ensemble is incorrect or incomplete.
In our studies we found that the correct free energy
is obtained using our method even when the histogram for
each coordinate was assumed to be flat, i.e., without the
use of a physical ensemble (data not shown).

\begin{figure}
    \includegraphics[scale=0.35]{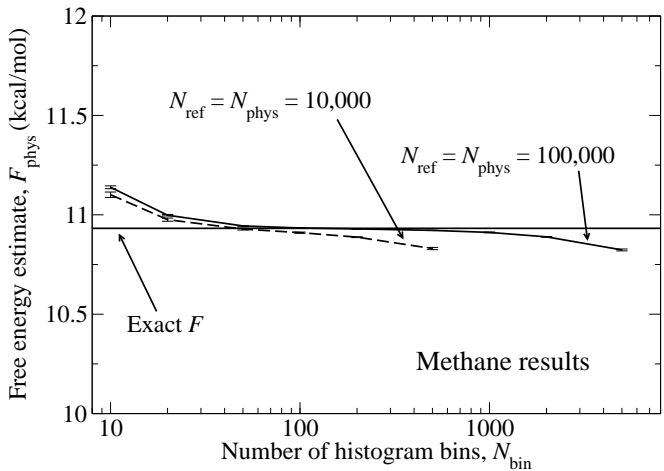}
    \caption{Absolute free energy for methane estimated by
	the reference system
	method as a function of the number of histogram bins used for
	each degree of freedom. The plot shows the ``sweet spot'' where
	histogram bins are small enough to reveal histogram features,
	yet large enough to give sufficient population in each bin.
	The results are shown with a vertical scale of
	two kcal/mol and on a log scale to emphasize the
	wide range of bin sizes that produce excellent results for the
	reference system approach.
	Results shown were obtained using Eq.\ (\ref{eq-fep})
	for a methane molecule using $N_{\rm phys}=N_{\rm ref}=10,000$
	(dashed curve)
	and $N_{\rm phys}=N_{\rm ref}=100,000$ (solid curve).
	The solid horizontal line shows the exact
	free energy and the errorbars are the standard deviations
	of five independent trials.
	The plot demonstrates at least fifty bins should
	be used for each independent coordinate,
	and that the maximum number of bins
	depends on the number of snapshots in the physical ensemble.
	\label{fig-sweet}
    }
\end{figure}

Choosing the size of the histogram bins
is an important consideration.
Figure \ref{fig-sweet} shows the large ``sweet spot'' where bins
are large enough
to be well populated, and yet small enough to reveal
histogram features.
The figure shows results for the absolute free energy
for a methane molecule using ten-thousand structures
in both the physical and reference ensembles,
$N_{\rm phys}=N_{\rm ref}=10,000$, (dashed curve)
and $N_{\rm phys}=N_{\rm ref}=100,000$ (solid curve).
The small vertical scale of two kcal/mol and the logarithmic horizontal
scale emphasize that there
is a wide range of bin sizes that produce excellent results for the
reference system approach.
Error bars are the standard deviation
of five independent simulations. The solid horizontal line shows the exact
free energy and the curves are free energy estimates,
using Eq.\  (\ref{eq-fep})
as a function of the number of bins used for the histograms
for all degrees of freedom. From this plot it is clear that one
should choose at least fifty bins, and that the maximum number of bins
that should be used depends on the number of snapshots in the physical
ensemble---more snapshots in the physical ensemble
means one can use more bins for the reference system.

\begin{table}
    \begin{tabular}{l|c|c}
    \hline \hline
    System & Estimate (kcal/mol) & Independent Estimate\\
    \hline
    $F_{\rm alpha}$ & 87.3 (0.7) & --- \\
    \hline
    $F_{\rm beta}$  & 86.3 (0.7) & --- \\
    \hline
    $\Delta F_{\rm alpha \rightarrow beta}$ & -1.0 (0.9) & -0.85 (0.05) \\
    \hline \hline
    \end{tabular}
    \caption{
	Absolute free energy estimates of
	the alpha ($F_{\rm alpha}$) and beta ($F_{\rm beta}$) conformations
	obtained using the 
	reference system method for leucine dipeptide with
	GBSA solvation, in units of kcal/mol.
	The independent measurement for the free energy difference
	was obtained via a 1.0 $\mu$s unconstrained simulation.
	The uncertainty for the absolute free energies, 
	shown in parentheses, is the standard deviation from five
	independent 10.0 ns leucine dipeptide simulations using
	one-million reference structures in the reference ensemble.
	The uncertainty
	for the free energy differences is obtained by using every possible
	combination of $F_{\rm alpha}$ and $F_{\rm beta}$,
	i.e., twenty-five independent estimates.
        The standard error associated with the 
	$\Delta F_{\rm alpha \rightarrow beta}$ reference system
	estimate is 0.18 kcal/mol, reflecting the twenty-five
	independent estimates.
	The table shows estimates of the configurational
	integral in Eq.\ (\ref{eq-Zphys}),
	i.e., the constant term is not included in the estimate.
    \label{tab-results2}
    }
\end{table}

\subsection{Leucine dipeptide}
Table \ref{tab-results2}
shows the agreement for leucine dipeptide
(ACE-(leu)$_2$-NME) between the free energy difference
$\Delta F_{\rm alpha \rightarrow beta}$
as predicted by the reference system method, and as
predicted via long simulation.
The leucine dipeptide physical ensembles were
generated using TINKER 4.2 \cite{tinker} with
the OPLS-AA forcefield. \cite{oplsaa}
The temperature was maintained at 500.0 K (to enable
an independent $\Delta F$ estimate via
repeated crossing of the free energy barrier between
alpha and beta configurations),
using Langevin dynamics with a friction coefficient of
5.0 ${\rm ps}^{-1}$. GBSA \cite{still} implicit
solvation was used, and RATTLE was utilized to maintain all bonds involving
hydrogens at their ideal lengths \cite{rattle} allowing the use
of a 2.0 fs time step.

We calculated reference systems and
computed absolute free energies of the alpha and
beta conformations based on five
10.0 ns trajectories. For all simulations, 
backbone torsions were constrained using a flat-bottomed
harmonic restraint (zero force if the torsion
angles were within the allowed range, and harmonic otherwise),
namely, for alpha: $-105<\phi<-45 \;{\rm and}\; -70<\psi<-10$;
and for beta: $-125<\phi<-65 \;{\rm and}\; 120<\psi<180$.
The reference system was generated using 100,000 snapshots
from the physical ensemble, then free energy estimates were obtained
by generating 1,000,000 structures for the reference ensemble for
each estimate. All one-hundred fifteen
(excludes bond lengths constrained by RATTLE \cite{rattle})
internal coordinates were binned as independent
with fifty bins for each coordinate.
The uncertainty shown in parenthesis is
the standard deviation from the
five independent estimates using the five separate trajectories, i.e.,
five different physical ensembles and five different reference systems.

Since independent estimates of the absolute free energies 
of the alpha and beta conformations of leucine dipeptide
are not available, we calculated the free
energy difference
$\Delta F_{\rm alpha \rightarrow beta} = -0.85 \; (0.05)$ kcal/mol
via a 1.0 $\mu$s unconstrained simulation.
The uncertainty of the independent estimate was obtained using
block averages.
The temperature was chosen to be 500.0 K which allowed around 1500
crossings of the free energy barrier between the alpha and
beta conformations, providing an accurate independent estimate.
As can be seen in Table \ref{tab-results2}, our estimated free
energy difference is in good agreement with the independent
value obtained via long simulation.

We emphasize that the nearly kcal/mol fluctuations observed in our
leucine dipeptide estimates are completely independent of the magnitude
of the free energy difference of the same order. That is, for a similar
sized system and similar CPU investment, one would expect similar uncertainty,
even for a very large free energy difference. This, indeed, is the motivation
for performing absolute free energy calculations. We believe, moreover, that
efficiency improvements will be achieved beyond the data in this initial
report.

\begin{figure}
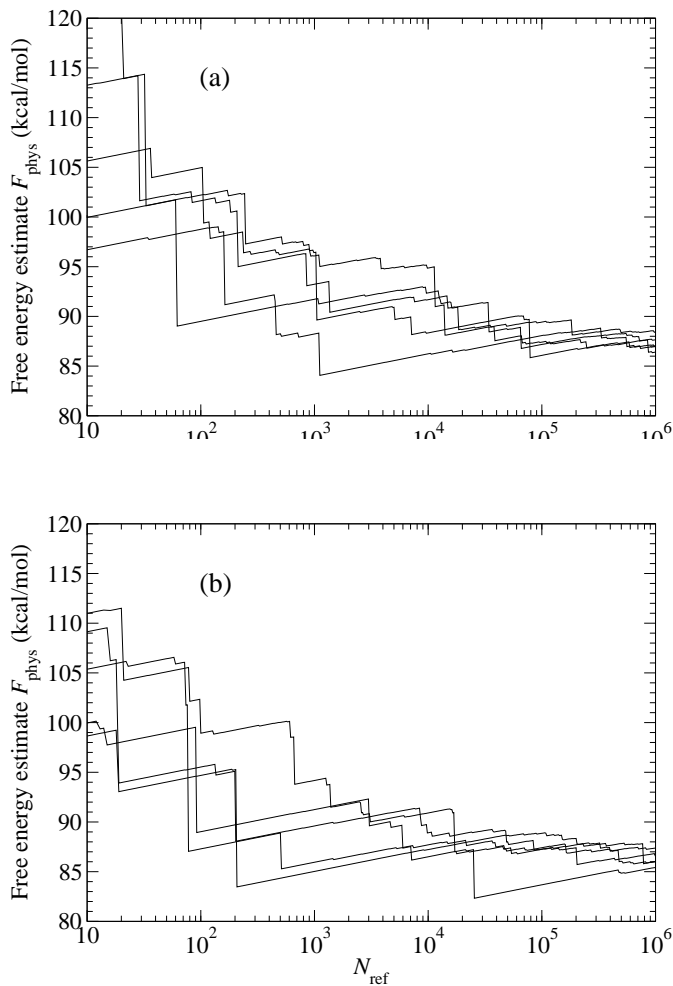

    \includegraphics[scale=0.35]{fig4.eps}\\
    \vspace{12pt}
    \includegraphics[scale=0.35]{fig5.eps}
    \caption{Free energy for leucine dipeptide estimated by
	the reference system
	method as a function of the number of reference structures
	$N_{\rm ref}$ used in the estimate.
	Five independent simulations are shown on a log scale to demonstrate
	the convergence behavior of the free energy estimate for
	(a) the alpha configuration, and (b) the beta configuration.
	Results shown were obtained using Eq.\ (\ref{eq-fep})
	with fifty bins for each degree of freedom.
	\label{fig-converge-di}
    }
\end{figure}

Figure \ref{fig-converge-di} shows the convergence
behavior of the reference
system method for leucine dipeptide. Five free energy
estimates are shown as a function of the number of reference
structures used in the estimate for
(a) the alpha configuration, and (b) the beta configuration.
Each of the five simulations use the same protocol as described above.

\begin{figure}
    \includegraphics[scale=0.35]{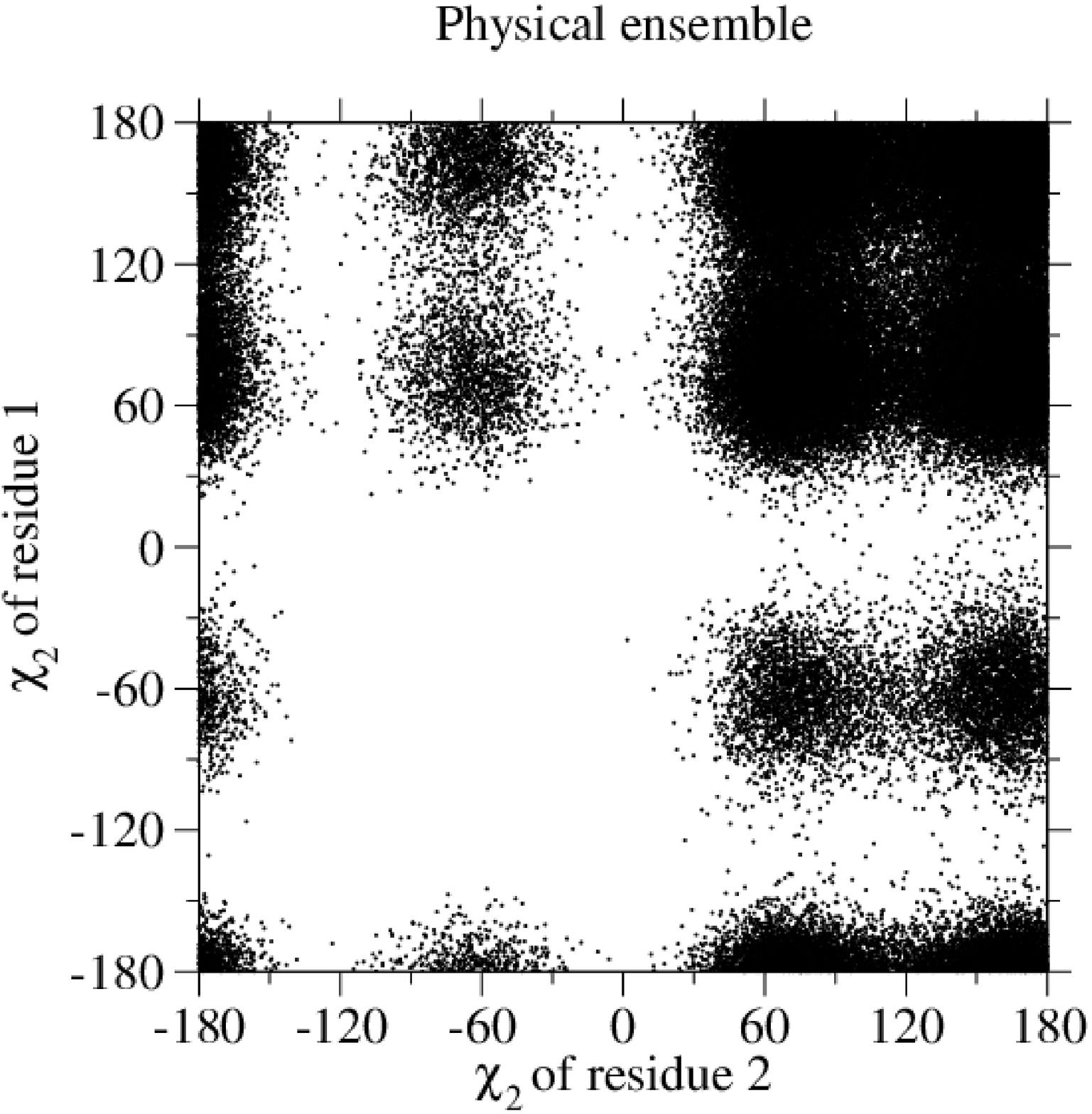}\\
    \vspace{12pt}
    \includegraphics[scale=0.35]{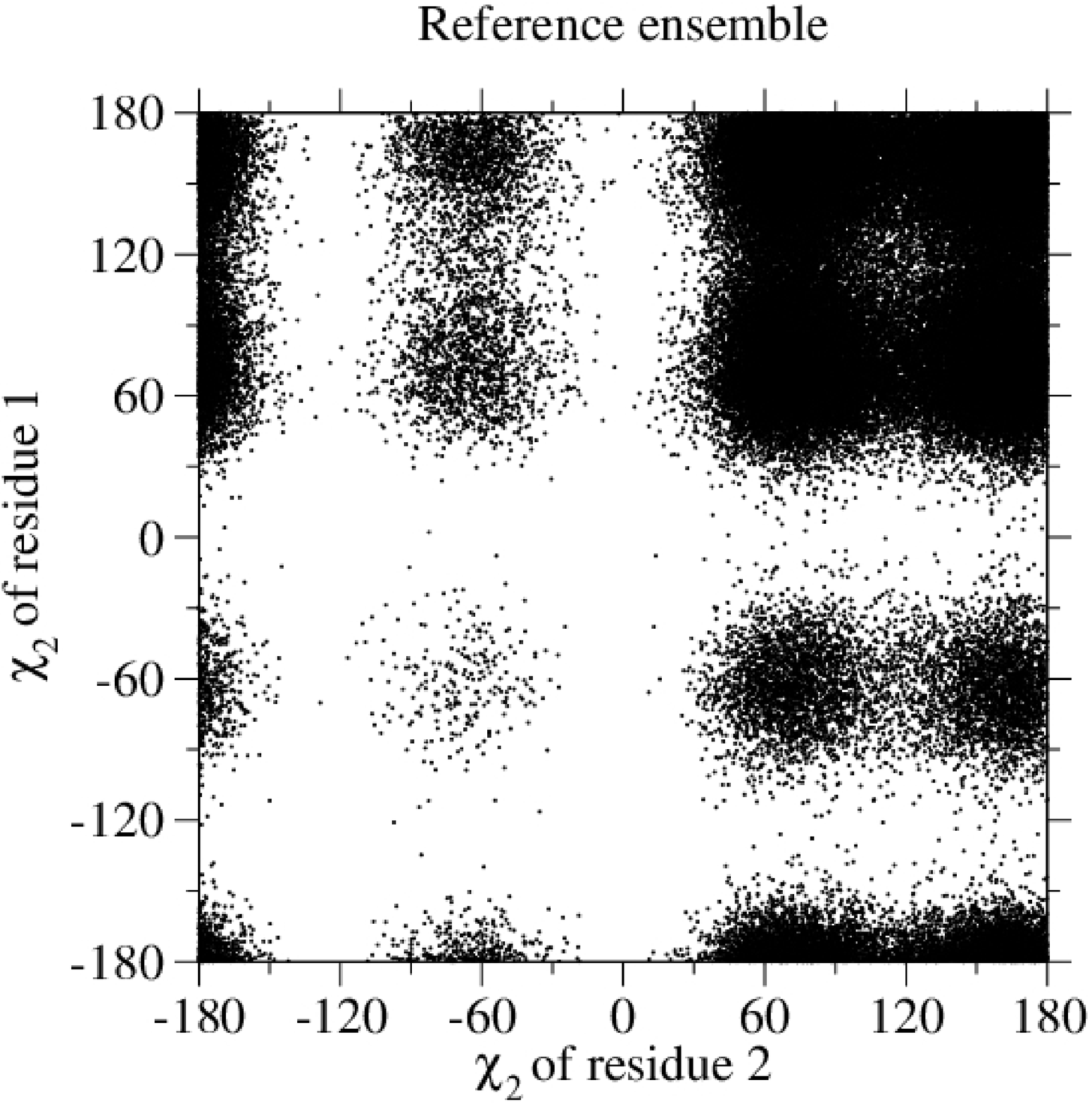}
    \caption{Scatter plots of the two $\chi_2$ torsions
	of each residue for leucine dipeptide. Results are shown
	for both physical and reference ensembles containing 100,000
	structures each.
	The figure shows that:
	(i) the reference system has good overlap with the physical system,
	as can be seen by the similarity between the two plots;
	and (ii) the reference system is more broadly distributed
	than the physical
	system, as evidenced by the data at (-60,-60) for the reference
	system that is not present for the physical system.
	\label{fig-scat}
    }
\end{figure}

The leucine dipeptide calculations also demonstrate two important
aspects of the particular reference system defined in this study:
(i) the reference system has good overlap with the physical system; and
(ii) the reference system is broader than the physical system.
Figure \ref{fig-scat} shows a scatter plot of the
$\chi_2$ torsions of each residue
for both the physical and reference ensembles. Each ensemble
contains 100,000 structures. The figure clearly shows the
excellent overlap between the reference and physical ensemble,
as can be seen by the similarity between the two plots. In
addition, the reference ensemble scatter plot has data
in the region (-60,-60) which does not exist in the
physical ensemble, showing that the reference system is ``broader'' than
the physical system.

\begin{figure}
    \includegraphics[scale=0.35]{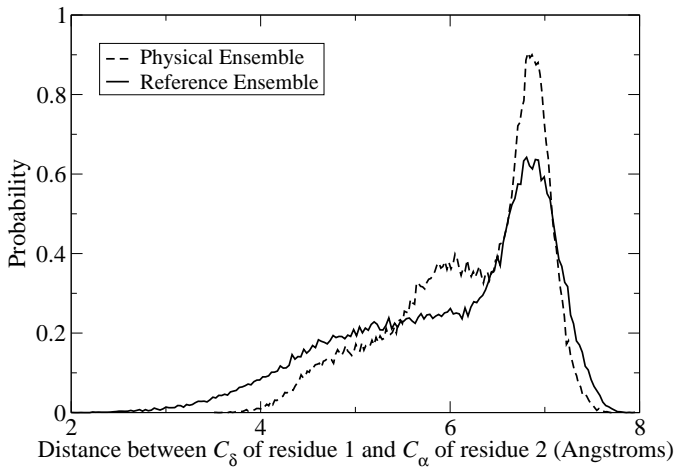}
    \caption{Histogram of the distance between the $C_\delta$ of residue one
	and the $C_\alpha$ of residue two for leucine dipeptide. Results are
	shown for both reference and physical ensembles containing 100,000
	structures each.
	The figure shows that:
	(i) the reference system has good overlap with the physical system;
	and (ii) the reference system is broader than the physical system.
	\label{fig-dist}
    }
\end{figure}

Figure \ref{fig-dist} shows a histogram of the distance between the $C_\delta$
atom of residue one and the $C_\alpha$ of residue two for 
the same ensembles as Fig.\ \ref{fig-scat}. The figure again shows
how the reference system has both excellent overlap with the
physical system and is also broader than the physical system.

\section{Discussion}
The present results raise a number of questions regarding the reference
system approach to computing absolute free energies---in particular, regarding
the use of correlations, the importance of the physical ensemble,
and the potential for application to larger systems.

\subsection{Correlation of Coordinates}
How can correlations among coordinates be used to increase the method's
effectiveness? One may choose to 
bin coordinates as independent (i.e., one-dimensional
histograms), or with correlations
(i.e., multi-dimensional histograms).
For example, in peptides, one may choose to bin all
sets of backbone $\phi,\psi$ torsions as correlated, and all other
coordinates (bond lengths, bond angles, other torsions) as
independent. It might always seem advantageous
to bin some coordinates (at least backbone torsions)
as correlated, since reference structures drawn
randomly from the histograms 
will be less likely to have steric
clashes. On the other hand, including correlations with small bin
sizes is impractical. As an example, imagine that for the leucine
dipeptide molecule used in
this study, one binned the four $\phi,\psi$ backbone torsions as
correlated. If fifty bins for each torsion were used (as should
be done according to the discussion below), then there
would be $50^4=6,250,000$ multi-dimensional bins to populate,
which is simply not feasible.

There does appear to be an important advantage to eliminating at
least some correlations from the original ``physical'' ensemble:
namely, a larger portion of conformational space
becomes available to the reference ensemble;
see Figs.\ \ref{fig-scat} and \ref{fig-dist}. 
Since coordinates for the reference structures
are drawn randomly and independently, it is
possible to generate reference structures that are
in entirely different energy basins than those in
the physical ensemble. {\it It is thus possible to
overcome the inadequacies of the physical ensemble
by binning internal coordinates independently}.
The optimal (presumably) limited use of correlations
will be considered in future work.

Regardless of the degree of correlations included in $U_{\rm ref}$,
we emphasize that final results fully include correlations in the physical
potential $U_{\rm phys}$.

\subsection{Quality of the physical ensemble}
Since the reference ensemble is generated by drawing at random from
histograms which, in turn, were generated from the physical ensemble,
a natural question to ask is: how complete does the physical
ensemble need to be?
The surprising answer is that, for our reference system method,
the physical ensemble does not need to
be complete, or even correct (properly distributed).
Since Eqs.\ (\ref{eq-Fphys}) and (\ref{eq-Fphys2}) are
valid for arbitrary reference systems,
the convergence of the free energy estimate to the correct value
is guaranteed, in the limit of infinite sampling
($N_{\rm ref} \rightarrow \infty$), regardless of the
quality of the physical ensemble.
The ``trick'' is that the ensemble for the reference system must
be converged, which can be achieved with much less expense since
there is no dynamical trapping.
Unlike the typical case for molecular mechanics simulation,
we sample the reference ensemble ``perfectly''---there is no possibility
of being trapped in a local basin. By construction, since all coordinate
values are generated exactly according to the reference distributions,
the reference ensemble can only suffer from statistical (but not systematic)
error.
For example, it was possible to obtain the correct
free energy for methane based on 10,000 reference structures
even when the histogram for
each coordinate was assumed to be flat, i.e., without
the use of a physical ensemble (data not shown).

It is important to note that, while convergence to the
correct free energy is guaranteed for any
choice of reference system, the efficiency of the method could
be dramatically reduced if the reference system does not overlap
well with the physical system.

Given the fact that the physical ensemble need not be correct, it
is easy to imagine a modified method that does not require
simulation, but instead populates the histogram bins using the ``bare''
potential for each internal coordinate (e.g., Gaussian histograms
for bond lengths and angles). Of course,
the conformational state must be defined explicitly,
with upper and lower limits for coordinates.
Allowed ranges for the torsions (especially
$\phi,\psi$) are naturally obtainable via, e.g., Ramachandran
propensities (e.g., Ref.\ \onlinecite{richardson}),
and reasonable ranges for bond lengths and angles
could be chosen to be, e.g., several standard deviations
from the mean.

\subsection{Extension to larger systems}
While the initial results of our reference system method are
promising, a naive implementation of the
method will find difficulty with large systems (as do
all absolute and relative free energy methods).
For our method, the difficulty
with including a very large number of degrees of freedom
is due to the fact that,
if one does not treat all correlations in
the backbone, then steric clashes will occur frequently when
generating the reference ensemble.

However, it is possible to extend the method
to larger peptides, still include all degrees of freedom, and
bin all coordinates independently (important for broadening
configurational space, as discussed above), by using
a ``segmentation'' technique motivated by earlier work.
\cite{gibson-seg,leach-seg}
Consider generating reference
structures for a ten-residue peptide in the alpha helix conformation.
Due to the large number of backbone torsions,
most of the reference structures chosen at random
will not be energetically favorable.
However, if one breaks the peptide into two pieces, then one can
generate many structures for each segment, and only
``keep'' energetically likely segment structures.
The selected structures
may be joined to form full structures which are reasonably likely
to have low energy. 
For example, if one generates $10^5$ structures for each of the
two segments and keeps only $10^3$ of those, then one only need
evaluate $10^3 \times 10^3 = 10^6$ full structures
out of a possible $10^5 \times 10^5 = 10^{10}$.
A statistically correct segmentation strategy
is currently being investigated by the authors for use in
large peptides.

Another strategy which may prove useful for larger systems
is to use the reference system method with multi-stage simulation.
Multi-stage
simulation requires the introduction of a hybrid potential energy
parameterized by $\lambda$, e.g.,
\begin{eqnarray}
    U_{\lambda} = \lambda U_{\rm phys} + (1-\lambda) U_{\rm ref}.
\end{eqnarray}
Thus, $U_0 = U_{\rm ref}$ and $U_1 = U_{\rm phys}$.
Simulations are performed using the hybrid potential energy
$U_{\lambda}$ (and thus a hybrid forcefield, if using molecular
dynamics) at intermediate $\lambda$ values between 0 and 1.
Conventional free energy methods such as thermodynamic integration
or free energy perturbation can then be used to
obtain $F_{\rm phys}$.

We also believe that including correlations, such as suggested
by Eq.\ (\ref{eq-Pcorr}) and possibly other ways, may be useful.
The inclusion of correlations should improve the overlap between
the reference and physical ensembles---thereby reducing the amount
of sampling required in the reference system, hence improving efficiency.
This also will be explored in future work.
(We also remind the
reader that the final free energy value includes the full correlations
in $U_{\rm phys}$, regardless of $U_{\rm ref}$.)

The method could prove useful in future protein-ligand binding
studies. In the simplest approach, one could freeze all degrees of
freedom except for the ligand and side-chain degrees of freedom
in the binding site. While the absolute free energy would be unphysical,
the approach could permit comparison of ligands or protein mutations
with little or no conformational similarity.

In principle, it is possible to extend the reference
system method to include explicitly solvated biomolecules.
However, as with all absolute free energy methods, the
addition of the solvent degrees of freedom causes
the free energy estimate to converge much more slowly than
without explicit solvent.
Thus, we feel the method described in this study will find
use primarily in implicitly solvated biomolecules.

\section{Conclusions}
In conclusion, we have introduced and tested a simple
method for calculating absolute
free energies in molecular systems.
The approach relies on the construction of an ensemble of
reference structures (i.e., the reference system) 
that is designed to have high overlap with the physical system
of interest.
The method was first shown to reproduce exactly computable
absolute free energies for simple systems, and
then used to correctly predict the stability of leucine
dipeptide conformations
using all one-hundred fifteen degrees of freedom.

Some strengths of the approach are that:
(i) the reference system is built to have good overlap with the system
of interest by using internal coordinates and by using
a single equilibrium ensemble from Monte Carlo or molecular dynamics;
(ii) the absolute free energy estimate is guaranteed to converge to the
correct value, whether or not the physical ensemble is complete
and, in fact, it is possible to estimate the absolute free energy
without the use of a physical ensemble;
(iii) the method explicitly includes all degrees of freedom employed
in the simulation;
(iv) the reference system need only be numerically
computable, i.e., the exact analytic result is not needed; and
(v) the method can  be trivially extended to include the use
of multi-stage simulation.
The CPU cost of the approach, beyond that for short trajectories
of the physical system of interest,
is one energy call for each reference structure, plus
the less expensive cost of generating the reference ensemble.

In the present ``proof of principle'' report,
our method was used to study conformational
equilibria; however we feel that the simplicity and flexibility
of the method may find broad use in computational biophysics
and biochemistry for a wide variety of free energy problems.
We have also described a segmentation strategy, currently being
pursued, to use the approach in much larger systems.

\section*{Acknowledgments}
The authors would like to thank Edward Lyman, Ronald White,
Srinath Cheluvarajah and Hagai Meirovitch for many
fruitful discussions.

\bibliography{}

\begin{thebibliography}{45}
\expandafter\ifx\csname natexlab\endcsname\relax\def\natexlab#1{#1}\fi
\expandafter\ifx\csname bibnamefont\endcsname\relax
  \def\bibnamefont#1{#1}\fi
\expandafter\ifx\csname bibfnamefont\endcsname\relax
  \def\bibfnamefont#1{#1}\fi
\expandafter\ifx\csname citenamefont\endcsname\relax
  \def\citenamefont#1{#1}\fi
\expandafter\ifx\csname url\endcsname\relax
  \def\url#1{\texttt{#1}}\fi
\expandafter\ifx\csname urlprefix\endcsname\relax\def\urlprefix{URL }\fi
\providecommand{\bibinfo}[2]{#2}
\providecommand{\eprint}[2][]{\url{#2}}

\bibitem[{\citenamefont{Grossfield et~al.}(2003)\citenamefont{Grossfield, Ren,
  and Ponder}}]{grossfield-jacs}
\bibinfo{author}{\bibfnamefont{A.}~\bibnamefont{Grossfield}},
  \bibinfo{author}{\bibfnamefont{P.}~\bibnamefont{Ren}}, \bibnamefont{and}
  \bibinfo{author}{\bibfnamefont{J.~W.} \bibnamefont{Ponder}},
  \bibinfo{journal}{J.\ Am.\ Chem.\ Soc.} \textbf{\bibinfo{volume}{125}},
  \bibinfo{pages}{15671} (\bibinfo{year}{2003}).

\bibitem[{\citenamefont{Pitera and van Gunsteren}(2001)}]{vangunsteren-onestep}
\bibinfo{author}{\bibfnamefont{J.~W.} \bibnamefont{Pitera}} \bibnamefont{and}
  \bibinfo{author}{\bibfnamefont{W.~F.} \bibnamefont{van Gunsteren}},
  \bibinfo{journal}{J.\ Phys.\ Chem.\ B} \textbf{\bibinfo{volume}{105}},
  \bibinfo{pages}{11264} (\bibinfo{year}{2001}).

\bibitem[{\citenamefont{Singh et~al.}(1994)\citenamefont{Singh, Ajay, Wemmer,
  and Kollman}}]{kollman-pnas}
\bibinfo{author}{\bibfnamefont{S.~B.} \bibnamefont{Singh}},
  \bibinfo{author}{\bibnamefont{Ajay}}, \bibinfo{author}{\bibfnamefont{D.~E.}
  \bibnamefont{Wemmer}}, \bibnamefont{and}
  \bibinfo{author}{\bibfnamefont{P.~A.} \bibnamefont{Kollman}},
  \bibinfo{journal}{Proc.\ Nat.\ Acad.\ Sci.\ (USA)}
  \textbf{\bibinfo{volume}{91}}, \bibinfo{pages}{7673} (\bibinfo{year}{1994}).

\bibitem[{\citenamefont{Oostenbrink et~al.}(2000)\citenamefont{Oostenbrink,
  Pitera, van Lipzip, Meerman, and van Gunsteren}}]{vangunsteren-estrogen}
\bibinfo{author}{\bibfnamefont{B.~C.} \bibnamefont{Oostenbrink}},
  \bibinfo{author}{\bibfnamefont{J.~W.} \bibnamefont{Pitera}},
  \bibinfo{author}{\bibfnamefont{M.~M.} \bibnamefont{van Lipzip}},
  \bibinfo{author}{\bibfnamefont{J.~H.~N.} \bibnamefont{Meerman}},
  \bibnamefont{and} \bibinfo{author}{\bibfnamefont{W.~F.} \bibnamefont{van
  Gunsteren}}, \bibinfo{journal}{J.\ Med.\ Chem.}
  \textbf{\bibinfo{volume}{43}}, \bibinfo{pages}{4594} (\bibinfo{year}{2000}).

\bibitem[{\citenamefont{Ytreberg and Zuckerman}(2005)}]{ytreberg-shift}
\bibinfo{author}{\bibfnamefont{F.~M.} \bibnamefont{Ytreberg}} \bibnamefont{and}
  \bibinfo{author}{\bibfnamefont{D.~M.} \bibnamefont{Zuckerman}},
  \bibinfo{journal}{J.\ Phys.\ Chem.\ B} \textbf{\bibinfo{volume}{109}},
  \bibinfo{pages}{9096} (\bibinfo{year}{2005}).

\bibitem[{\citenamefont{Shoichet}(2004)}]{shoichet-nature}
\bibinfo{author}{\bibfnamefont{B.~K.} \bibnamefont{Shoichet}},
  \bibinfo{journal}{Nature} \textbf{\bibinfo{volume}{432}},
  \bibinfo{pages}{862} (\bibinfo{year}{2004}).

\bibitem[{\citenamefont{Trosset and Scheraga}(1999)}]{scheraga}
\bibinfo{author}{\bibfnamefont{J.~Y.} \bibnamefont{Trosset}} \bibnamefont{and}
  \bibinfo{author}{\bibfnamefont{H.~A.} \bibnamefont{Scheraga}},
  \bibinfo{journal}{J.\ Comput.\ Chem.} \textbf{\bibinfo{volume}{20}},
  \bibinfo{pages}{412} (\bibinfo{year}{1999}).

\bibitem[{\citenamefont{Zwanzig}(1954)}]{zwanzig}
\bibinfo{author}{\bibfnamefont{R.~W.} \bibnamefont{Zwanzig}},
  \bibinfo{journal}{J.\ Chem.\ Phys.} \textbf{\bibinfo{volume}{22}},
  \bibinfo{pages}{1420} (\bibinfo{year}{1954}).

\bibitem[{\citenamefont{Beveridge and Di{C}apua}(1989)}]{beveridge}
\bibinfo{author}{\bibfnamefont{D.}~\bibnamefont{Beveridge}} \bibnamefont{and}
  \bibinfo{author}{\bibfnamefont{F.}~\bibnamefont{Di{C}apua}},
  \bibinfo{journal}{Ann.\ Rev.\ Biophys.\ Biophys.\ Chem.}
  \textbf{\bibinfo{volume}{18}}, \bibinfo{pages}{431} (\bibinfo{year}{1989}).

\bibitem[{\citenamefont{Jorgensen and Ravimohan}(1985)}]{jorgensen}
\bibinfo{author}{\bibfnamefont{W.~L.} \bibnamefont{Jorgensen}}
  \bibnamefont{and}
  \bibinfo{author}{\bibfnamefont{C.}~\bibnamefont{Ravimohan}},
  \bibinfo{journal}{J.\ Chem.\ Phys.} \textbf{\bibinfo{volume}{83}},
  \bibinfo{pages}{3050} (\bibinfo{year}{1985}).

\bibitem[{\citenamefont{Yang et~al.}(2004)\citenamefont{Yang, Bitetti-Putzer,
  and Karplus}}]{karplus-jcp}
\bibinfo{author}{\bibfnamefont{W.}~\bibnamefont{Yang}},
  \bibinfo{author}{\bibfnamefont{R.}~\bibnamefont{Bitetti-Putzer}},
  \bibnamefont{and} \bibinfo{author}{\bibfnamefont{M.}~\bibnamefont{Karplus}},
  \bibinfo{journal}{J.\ Chem.\ Phys.} \textbf{\bibinfo{volume}{120}},
  \bibinfo{pages}{2618} (\bibinfo{year}{2004}).

\bibitem[{\citenamefont{Mc{C}ammon}(1991)}]{mccammon}
\bibinfo{author}{\bibfnamefont{J.~A.} \bibnamefont{Mc{C}ammon}},
  \bibinfo{journal}{Curr.\ Opin.\ Struc.\ Bio.} \textbf{\bibinfo{volume}{2}},
  \bibinfo{pages}{96} (\bibinfo{year}{1991}).

\bibitem[{\citenamefont{Hoover et~al.}(1971)\citenamefont{Hoover, Gray, and
  Johnson}}]{hoover71}
\bibinfo{author}{\bibfnamefont{W.~G.} \bibnamefont{Hoover}},
  \bibinfo{author}{\bibfnamefont{S.~G.} \bibnamefont{Gray}}, \bibnamefont{and}
  \bibinfo{author}{\bibfnamefont{K.~W.} \bibnamefont{Johnson}},
  \bibinfo{journal}{J.\ Chem.\ Phys.} \textbf{\bibinfo{volume}{55}},
  \bibinfo{pages}{1128} (\bibinfo{year}{1971}).

\bibitem[{\citenamefont{Frenkel and Ladd}(1984)}]{frenkel}
\bibinfo{author}{\bibfnamefont{D.}~\bibnamefont{Frenkel}} \bibnamefont{and}
  \bibinfo{author}{\bibfnamefont{A.~J.~C.} \bibnamefont{Ladd}},
  \bibinfo{journal}{J.\ Chem.\ Phys.} \textbf{\bibinfo{volume}{81}},
  \bibinfo{pages}{3188} (\bibinfo{year}{1984}).

\bibitem[{\citenamefont{Hoover and Ree}(1967)}]{hoover67}
\bibinfo{author}{\bibfnamefont{W.~G.} \bibnamefont{Hoover}} \bibnamefont{and}
  \bibinfo{author}{\bibfnamefont{F.~H.} \bibnamefont{Ree}},
  \bibinfo{journal}{J.\ Chem.\ Phys.} \textbf{\bibinfo{volume}{47}},
  \bibinfo{pages}{4873} (\bibinfo{year}{1967}).

\bibitem[{\citenamefont{Amon and Reinhardt}(2000)}]{reinhardt-absf}
\bibinfo{author}{\bibfnamefont{L.~M.} \bibnamefont{Amon}} \bibnamefont{and}
  \bibinfo{author}{\bibfnamefont{W.~P.} \bibnamefont{Reinhardt}},
  \bibinfo{journal}{J.\ Chem.\ Phys.} \textbf{\bibinfo{volume}{113}},
  \bibinfo{pages}{3573} (\bibinfo{year}{2000}).

\bibitem[{\citenamefont{Stoessel and Nowak}(1990)}]{stoessel}
\bibinfo{author}{\bibfnamefont{J.~P.} \bibnamefont{Stoessel}} \bibnamefont{and}
  \bibinfo{author}{\bibfnamefont{P.}~\bibnamefont{Nowak}},
  \bibinfo{journal}{Macromolecules} \textbf{\bibinfo{volume}{23}},
  \bibinfo{pages}{1961} (\bibinfo{year}{1990}).

\bibitem[{\citenamefont{Cheluvaraja and
  Meirovitch}(2004{\natexlab{a}})}]{meirovitch-deca}
\bibinfo{author}{\bibfnamefont{S.}~\bibnamefont{Cheluvaraja}} \bibnamefont{and}
  \bibinfo{author}{\bibfnamefont{H.}~\bibnamefont{Meirovitch}},
  \bibinfo{journal}{Proc.\ Nat.\ Acad.\ Sci.} \textbf{\bibinfo{volume}{101}},
  \bibinfo{pages}{9241} (\bibinfo{year}{2004}{\natexlab{a}}).

\bibitem[{\citenamefont{White and Meirovitch}(2004)}]{meirovitch-argon}
\bibinfo{author}{\bibfnamefont{R.~P.} \bibnamefont{White}} \bibnamefont{and}
  \bibinfo{author}{\bibfnamefont{H.}~\bibnamefont{Meirovitch}},
  \bibinfo{journal}{Proc.\ Nat.\ Acad.\ Sci.} \textbf{\bibinfo{volume}{101}},
  \bibinfo{pages}{9235} (\bibinfo{year}{2004}).

\bibitem[{\citenamefont{Cheluvaraja and
  Meirovitch}(2004{\natexlab{b}})}]{meirovitch-jcp}
\bibinfo{author}{\bibfnamefont{S.}~\bibnamefont{Cheluvaraja}} \bibnamefont{and}
  \bibinfo{author}{\bibfnamefont{H.}~\bibnamefont{Meirovitch}},
  \bibinfo{journal}{J.\ Chem.\ Phys.} \textbf{\bibinfo{volume}{1022}},
  \bibinfo{pages}{054903} (\bibinfo{year}{2004}{\natexlab{b}}).

\bibitem[{\citenamefont{Head et~al.}(1997)\citenamefont{Head, Given, , and
  Gilson}}]{gilson-jpca}
\bibinfo{author}{\bibfnamefont{M.~S.} \bibnamefont{Head}},
  \bibinfo{author}{\bibfnamefont{J.~A.} \bibnamefont{Given}}, ,
  \bibnamefont{and} \bibinfo{author}{\bibfnamefont{M.~K.}
  \bibnamefont{Gilson}}, \bibinfo{journal}{J.\ Phys.\ Chem.\ A}
  \textbf{\bibinfo{volume}{101}}, \bibinfo{pages}{1609} (\bibinfo{year}{1997}).

\bibitem[{\citenamefont{Gilson et~al.}(1997)\citenamefont{Gilson, Given, Bush,
  and Mc{C}ammon}}]{gilson-bj}
\bibinfo{author}{\bibfnamefont{M.~K.} \bibnamefont{Gilson}},
  \bibinfo{author}{\bibfnamefont{J.~A.} \bibnamefont{Given}},
  \bibinfo{author}{\bibfnamefont{B.~L.} \bibnamefont{Bush}}, \bibnamefont{and}
  \bibinfo{author}{\bibfnamefont{J.~A.} \bibnamefont{Mc{C}ammon}},
  \bibinfo{journal}{Biophys.\ J.} \textbf{\bibinfo{volume}{72}},
  \bibinfo{pages}{1047} (\bibinfo{year}{1997}).

\bibitem[{\citenamefont{Chang and Gilson}(2004)}]{gilson-jacs}
\bibinfo{author}{\bibfnamefont{C.~E.} \bibnamefont{Chang}} \bibnamefont{and}
  \bibinfo{author}{\bibfnamefont{M.~K.} \bibnamefont{Gilson}},
  \bibinfo{journal}{J.\ Am.\ Chem.\ Soc.} \textbf{\bibinfo{volume}{126}},
  \bibinfo{pages}{13156} (\bibinfo{year}{2004}).

\bibitem[{\citenamefont{Karplus and Kushick}(1981)}]{karplus-deca}
\bibinfo{author}{\bibfnamefont{M.}~\bibnamefont{Karplus}} \bibnamefont{and}
  \bibinfo{author}{\bibfnamefont{J.~N.} \bibnamefont{Kushick}},
  \bibinfo{journal}{Macromolecules} \textbf{\bibinfo{volume}{14}},
  \bibinfo{pages}{325} (\bibinfo{year}{1981}).

\bibitem[{\citenamefont{Carlsson and Aqvist}(2005)}]{aqvist-absf}
\bibinfo{author}{\bibfnamefont{J.}~\bibnamefont{Carlsson}} \bibnamefont{and}
  \bibinfo{author}{\bibfnamefont{J.}~\bibnamefont{Aqvist}},
  \bibinfo{journal}{J.\ Phys.\ Chem.\ B} \textbf{\bibinfo{volume}{109}},
  \bibinfo{pages}{6448} (\bibinfo{year}{2005}).

\bibitem[{\citenamefont{Chang et~al.}(2003)\citenamefont{Chang, Potter, and
  Gilson}}]{gilson-jpcb}
\bibinfo{author}{\bibfnamefont{C.~E.} \bibnamefont{Chang}},
  \bibinfo{author}{\bibfnamefont{M.~J.} \bibnamefont{Potter}},
  \bibnamefont{and} \bibinfo{author}{\bibfnamefont{M.~K.}
  \bibnamefont{Gilson}}, \bibinfo{journal}{J.\ Phys.\ Chem.\ B}
  \textbf{\bibinfo{volume}{107}}, \bibinfo{pages}{1048} (\bibinfo{year}{2003}).

\bibitem[{\citenamefont{Bennett}(1976)}]{bennett}
\bibinfo{author}{\bibfnamefont{C.~H.} \bibnamefont{Bennett}},
  \bibinfo{journal}{J.\ Comput.\ Phys.} \textbf{\bibinfo{volume}{22}},
  \bibinfo{pages}{245} (\bibinfo{year}{1976}).

\bibitem[{\citenamefont{Shirts and Pande}(2005)}]{shirts-benn}
\bibinfo{author}{\bibfnamefont{M.~R.} \bibnamefont{Shirts}} \bibnamefont{and}
  \bibinfo{author}{\bibfnamefont{V.~S.} \bibnamefont{Pande}},
  \bibinfo{journal}{J.\ Chem.\ Phys.} \textbf{\bibinfo{volume}{122}},
  \bibinfo{pages}{144107} (\bibinfo{year}{2005}).

\bibitem[{\citenamefont{Shirts et~al.}(2003)\citenamefont{Shirts, Bair, Hooker,
  and Pande}}]{shirts-prl}
\bibinfo{author}{\bibfnamefont{M.~R.} \bibnamefont{Shirts}},
  \bibinfo{author}{\bibfnamefont{E.}~\bibnamefont{Bair}},
  \bibinfo{author}{\bibfnamefont{G.}~\bibnamefont{Hooker}}, \bibnamefont{and}
  \bibinfo{author}{\bibfnamefont{V.~S.} \bibnamefont{Pande}},
  \bibinfo{journal}{Phys.\ Rev.\ Lett.} \textbf{\bibinfo{volume}{91}},
  \bibinfo{pages}{140601} (\bibinfo{year}{2003}).

\bibitem[{\citenamefont{Crooks}(2000)}]{crooks-pre}
\bibinfo{author}{\bibfnamefont{G.~E.} \bibnamefont{Crooks}},
  \bibinfo{journal}{Phys.\ Rev.\ E} \textbf{\bibinfo{volume}{61}},
  \bibinfo{pages}{2361} (\bibinfo{year}{2000}).

\bibitem[{\citenamefont{Lu et~al.}(2004)\citenamefont{Lu, Kofke, and
  Woolf}}]{lu-jcc}
\bibinfo{author}{\bibfnamefont{N.}~\bibnamefont{Lu}},
  \bibinfo{author}{\bibfnamefont{D.~A.} \bibnamefont{Kofke}}, \bibnamefont{and}
  \bibinfo{author}{\bibfnamefont{T.~B.} \bibnamefont{Woolf}},
  \bibinfo{journal}{J.\ Comput.\ Chem.} \textbf{\bibinfo{volume}{25}},
  \bibinfo{pages}{28} (\bibinfo{year}{2004}).

\bibitem[{\citenamefont{Zuckerman and
  Woolf}(2002{\natexlab{a}})}]{zuckerman-prl}
\bibinfo{author}{\bibfnamefont{D.~M.} \bibnamefont{Zuckerman}}
  \bibnamefont{and} \bibinfo{author}{\bibfnamefont{T.~B.} \bibnamefont{Woolf}},
  \bibinfo{journal}{Phys.\ Rev.\ Lett.} \textbf{\bibinfo{volume}{89}},
  \bibinfo{pages}{180602} (\bibinfo{year}{2002}{\natexlab{a}}).

\bibitem[{\citenamefont{Zuckerman and Woolf}(2004)}]{zuckerman-jstat}
\bibinfo{author}{\bibfnamefont{D.~M.} \bibnamefont{Zuckerman}}
  \bibnamefont{and} \bibinfo{author}{\bibfnamefont{T.~B.} \bibnamefont{Woolf}},
  \bibinfo{journal}{J.\ Stat.\ Phys.} \textbf{\bibinfo{volume}{114}},
  \bibinfo{pages}{1303} (\bibinfo{year}{2004}).

\bibitem[{\citenamefont{Zuckerman and
  Woolf}(2002{\natexlab{b}})}]{zuckerman-cpl}
\bibinfo{author}{\bibfnamefont{D.~M.} \bibnamefont{Zuckerman}}
  \bibnamefont{and} \bibinfo{author}{\bibfnamefont{T.~B.} \bibnamefont{Woolf}},
  \bibinfo{journal}{Chem.\ Phys.\ Lett.} \textbf{\bibinfo{volume}{351}},
  \bibinfo{pages}{445} (\bibinfo{year}{2002}{\natexlab{b}}).

\bibitem[{\citenamefont{Ytreberg and
  Zuckerman}(2004{\natexlab{a}})}]{ytreberg-extrap}
\bibinfo{author}{\bibfnamefont{F.~M.} \bibnamefont{Ytreberg}} \bibnamefont{and}
  \bibinfo{author}{\bibfnamefont{D.~M.} \bibnamefont{Zuckerman}},
  \bibinfo{journal}{J.\ Comput.\ Chem.} \textbf{\bibinfo{volume}{25}},
  \bibinfo{pages}{1749} (\bibinfo{year}{2004}{\natexlab{a}}).

\bibitem[{\citenamefont{Straatsma and Mc{C}ammon}(1991)}]{straatsma-ti}
\bibinfo{author}{\bibfnamefont{T.~P.} \bibnamefont{Straatsma}}
  \bibnamefont{and} \bibinfo{author}{\bibfnamefont{J.~A.}
  \bibnamefont{Mc{C}ammon}}, \bibinfo{journal}{J.\ Chem.\ Phys.}
  \textbf{\bibinfo{volume}{95}}, \bibinfo{pages}{1175} (\bibinfo{year}{1991}).

\bibitem[{\citenamefont{Fasnacht et~al.}(2004)\citenamefont{Fasnacht, Swendsen,
  and Rosenberg}}]{swendsen-aim}
\bibinfo{author}{\bibfnamefont{M.}~\bibnamefont{Fasnacht}},
  \bibinfo{author}{\bibfnamefont{R.~H.} \bibnamefont{Swendsen}},
  \bibnamefont{and} \bibinfo{author}{\bibfnamefont{J.~M.}
  \bibnamefont{Rosenberg}}, \bibinfo{journal}{Phys.\ Rev.\ E}
  \textbf{\bibinfo{volume}{69}}, \bibinfo{pages}{056704}
  (\bibinfo{year}{2004}).

\bibitem[{\citenamefont{Andersen}(1983)}]{rattle}
\bibinfo{author}{\bibfnamefont{H.~C.} \bibnamefont{Andersen}},
  \bibinfo{journal}{J.\ Comput.\ Phys.} \textbf{\bibinfo{volume}{52}},
  \bibinfo{pages}{24} (\bibinfo{year}{1983}).

\bibitem[{\citenamefont{Ytreberg and
  Zuckerman}(2004{\natexlab{b}})}]{ytreberg-seps}
\bibinfo{author}{\bibfnamefont{F.~M.} \bibnamefont{Ytreberg}} \bibnamefont{and}
  \bibinfo{author}{\bibfnamefont{D.~M.} \bibnamefont{Zuckerman}},
  \bibinfo{journal}{J.\ Chem.\ Phys.} \textbf{\bibinfo{volume}{120}},
  \bibinfo{pages}{10876} (\bibinfo{year}{2004}{\natexlab{b}}).

\bibitem[{\citenamefont{Ponder and Richard}(1987)}]{tinker}
\bibinfo{author}{\bibfnamefont{J.~W.} \bibnamefont{Ponder}} \bibnamefont{and}
  \bibinfo{author}{\bibfnamefont{F.~M.} \bibnamefont{Richard}},
  \bibinfo{journal}{J.\ Comput.\ Chem.} \textbf{\bibinfo{volume}{8}},
  \bibinfo{pages}{1016} (\bibinfo{year}{1987}),
  \bibinfo{note}{http://dasher.wustl.edu/tinker}.

\bibitem[{\citenamefont{Jorgensen et~al.}(1996)\citenamefont{Jorgensen,
  Maxwell, and Tirado-Rives}}]{oplsaa}
\bibinfo{author}{\bibfnamefont{W.~L.} \bibnamefont{Jorgensen}},
  \bibinfo{author}{\bibfnamefont{D.~S.} \bibnamefont{Maxwell}},
  \bibnamefont{and}
  \bibinfo{author}{\bibfnamefont{J.}~\bibnamefont{Tirado-Rives}},
  \bibinfo{journal}{J.\ Am.\ Chem.\ Soc.} \textbf{\bibinfo{volume}{117}},
  \bibinfo{pages}{11225} (\bibinfo{year}{1996}).

\bibitem[{\citenamefont{Still et~al.}(1990)\citenamefont{Still, Tempczyk, and
  Hawley}}]{still}
\bibinfo{author}{\bibfnamefont{W.~C.} \bibnamefont{Still}},
  \bibinfo{author}{\bibfnamefont{A.}~\bibnamefont{Tempczyk}}, \bibnamefont{and}
  \bibinfo{author}{\bibfnamefont{R.~C.} \bibnamefont{Hawley}},
  \bibinfo{journal}{J.\ Am.\ Chem.\ Soc.} \textbf{\bibinfo{volume}{112}},
  \bibinfo{pages}{6127} (\bibinfo{year}{1990}).

\bibitem[{\citenamefont{Lovell et~al.}(2003)\citenamefont{Lovell, Davis,
  {Arendall III}, {de Bakker}, Word, Prisant, Richardson, and
  Richardson}}]{richardson}
\bibinfo{author}{\bibfnamefont{S.~C.} \bibnamefont{Lovell}},
  \bibinfo{author}{\bibfnamefont{I.~W.} \bibnamefont{Davis}},
  \bibinfo{author}{\bibfnamefont{W.~B.} \bibnamefont{{Arendall III}}},
  \bibinfo{author}{\bibfnamefont{P.~I.~W.} \bibnamefont{{de Bakker}}},
  \bibinfo{author}{\bibfnamefont{J.~M.} \bibnamefont{Word}},
  \bibinfo{author}{\bibfnamefont{M.~G.} \bibnamefont{Prisant}},
  \bibinfo{author}{\bibfnamefont{J.~S.} \bibnamefont{Richardson}},
  \bibnamefont{and} \bibinfo{author}{\bibfnamefont{D.~C.}
  \bibnamefont{Richardson}}, \bibinfo{journal}{Proteins}
  \textbf{\bibinfo{volume}{50}}, \bibinfo{pages}{437} (\bibinfo{year}{2003}).

\bibitem[{\citenamefont{Gibson and Scheraga}(1987)}]{gibson-seg}
\bibinfo{author}{\bibfnamefont{K.~D.} \bibnamefont{Gibson}} \bibnamefont{and}
  \bibinfo{author}{\bibfnamefont{H.~A.} \bibnamefont{Scheraga}},
  \bibinfo{journal}{J.\ Comput.\ Chem.} \textbf{\bibinfo{volume}{8}},
  \bibinfo{pages}{826} (\bibinfo{year}{1987}).

\bibitem[{\citenamefont{Leach et~al.}(1988)\citenamefont{Leach, Prout, and
  Dolata}}]{leach-seg}
\bibinfo{author}{\bibfnamefont{A.~R.} \bibnamefont{Leach}},
  \bibinfo{author}{\bibfnamefont{K.}~\bibnamefont{Prout}}, \bibnamefont{and}
  \bibinfo{author}{\bibfnamefont{D.~P.} \bibnamefont{Dolata}},
  \bibinfo{journal}{J.\ Comput.\ Aid.\ Mol.\ Des.}
  \textbf{\bibinfo{volume}{2}}, \bibinfo{pages}{107} (\bibinfo{year}{1988}).

\end{thebibliography}

\end{document}